\documentclass[journal,10pt,xcdraw,table]{IEEEtran}

\usepackage{cite}
\usepackage{amsmath,amssymb,amsfonts,bm}
\usepackage{algorithmic}
\usepackage{blindtext,graphicx}
\usepackage{textcomp}
\usepackage[dvipsnames]{xcolor}
\usepackage{caption}
\usepackage{subfloat}
\usepackage{comment}
\usepackage{courier}
\usepackage{multicol}
\usepackage{multirow}
\usepackage{pdfpages}
\usepackage{url}
\usepackage{hyperref}
\usepackage{cuted}

\renewcommand*{\\}{\vspace*{0.2cm}}
\usepackage{tikz}
\usetikzlibrary{math}
\definecolor{myred}{rgb}{0.77, 0.12, 0.23}

\ifCLASSINFOpdf
\else
\fi

\newcommand{\rev}[1]{\textcolor{black}{#1}}
\newcommand{\revtwo}[1]{\textcolor{black}{#1}}

\begin{document}
\cleardoublepage
\title{\huge{Computation of Optical Refractive Index Structure Parameter from its Statistical Definition Using Radiosonde Data}}

\author{\IEEEauthorblockN{Florian Quatresooz\IEEEauthorrefmark{1}, Danielle Vanhoenacker-Janvier\IEEEauthorrefmark{3}, Claude Oestges\IEEEauthorrefmark{4}}
\IEEEauthorblockA{\\ICTEAM - Universit\'e catholique de Louvain - Louvain-la-Neuve, Belgium\\}
\IEEEauthorrefmark{1}florian.quatresooz@uclouvain.be, \IEEEauthorrefmark{3}danielle.vanhoenacker@uclouvain.be, \IEEEauthorrefmark{4}claude.oestges@uclouvain.be
\vspace{-0.9cm}}
%\thanks{Manuscript received ...}}

%to add author's version on top of first page

\markboth{A\MakeLowercase{uthor's version.} A\MakeLowercase{n edited version of this paper was published by} AGU. C\MakeLowercase{opyright} 2023 A\MakeLowercase{merican} G\MakeLowercase{eophysical} U\MakeLowercase{nion}. C\MakeLowercase{itation information:} DOI \href{http://doi.org/10.1029/2022RS007624}{10.1029/2022RS007624}}{}

\maketitle

\begin{abstract}
Knowledge of the optical refractive index structure parameter $C_n^2$ is of interest for Free Space Optics (FSO) and ground-based optical astronomy, as it depicts the strength of the expected scintillation on the received optical waves. Focus is given here to models using meteorological quantities coming from radiosonde measurements as inputs to estimate the $C_n^2$ profile in the atmosphere. A model relying on the $C_n^2$ statistical definition is presented and applied to recent high-density radiosonde profiles at Trappes (France) and Hilo, HI (USA). It is also compared to thermosonde measurements coming from the T-REX campaign. This model enables to obtain site-specific average profiles and to identify isolated turbulent layers using only pressure and temperature measurements, paving the way for optical site selection. It offers similar performance when compared to a Tatarskii-based model inspired by the literature.
\end{abstract}

\IEEEpeerreviewmaketitle

\makeatletter
\def\ps@IEEEtitlepagestyle{
  \def\@oddfoot{\mycopyrightnotice}
  \def\@evenfoot{}
}
\def\mycopyrightnotice{
  {\footnotesize
  \begin{minipage}{\textwidth}
  \centering
  Copyright~\copyright~2023 American Geophysical Union.
  \end{minipage}
  }
}

\section{Introduction}
Earth-to-satellite optical communications and ground-based optical astronomy share a common interest: the characterization of atmospheric effects on the propagation of optical waves. \rev{Optical turbulence} is one of these effects and arises from random fluctuations of the refractive index due to atmospheric turbulence. This phenomenon is behind the twinkling of stars and the need for adaptive optics to correct optical observations. \rev{It has been studied by astronomers for more than 40 years and numerous models have been designed~\cite{roddier1981v}. However, one should be cautious when applying these models to optical communication sites since atmospheric conditions different than at astronomical sites are expected~\cite{de2018preliminary}. Hence, extra work on this topic is needed.}

Commonly, \rev{electro-optical} engineers are interested in the scintillation effects on the received electromagnetic waves, such as the scintillation index $\sigma_I^2$ or the variance of the log-amplitude fluctuations $\sigma_\chi^2$. Under some hypotheses, they directly depend on the refractive index structure parameter $C_n^2$ through analytical expressions \cite{tatarskii1971effects,ishimaru1978wave,andrews2005}. $C_n^2$ is a parameter that describes the intensity of atmospheric refractive index fluctuations and its knowledge enables to perform flexible scheduling or ground site selection (e.g. by avoiding hours or areas with large turbulence). It is also of particular interest for adaptive optics and for the design of optical communication systems. $C_n^2$ varies with time (hours of the day and seasons), as well as with space: it depends on the location on Earth and on the altitude. The altitude dependence of $C_n^2$ leads to $C_n^2$ profiles, which are interesting for ground-to-satellite communications crossing the whole atmosphere.

Measurements of $C_n^2$ profiles are not always practical since they require the use of dedicated tools (e.g.~thermosondes). Moreover, some tools can only provide integrated quantities of $C_n^2$ profiles, such as the seeing that can be obtained with a differential image motion monitor for example \cite{tokovinin2002differential}. Hence, throughout the years, several models have been designed to determine $C_n^2$ profiles based only on atmospheric quantities, namely using meteorological quantities. 

{\begin{figure}
    \centering
    \includegraphics[width=0.5\textwidth]{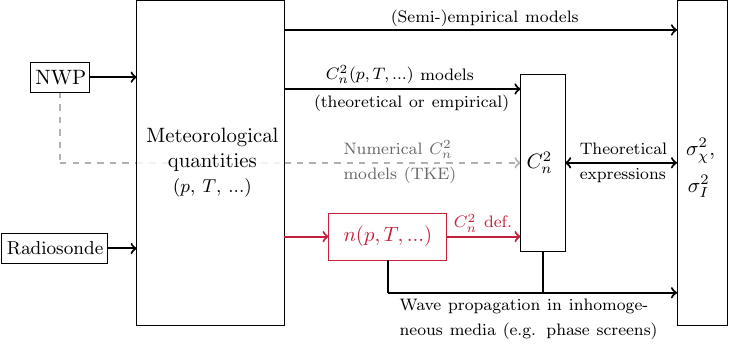}
    \caption{Main scintillation quantities and associated links.}
    \label{fig:schematic}
\end{figure}
}

Different approaches classically found in the literature are depicted in Fig.~\ref{fig:schematic}. Meteorological quantities can come from radiosonde measurements or from numerical weather prediction (NWP) simulations; the latter enabling the prediction of future $C_n^2$ profiles based on current weather forecast. From these quantities, empirical (e.g. Dewan model \cite{dewan1993model}, Hufnagel-Valley model \cite{andrews2005,HVchapter6}) or theoretical (e.g. Tatarskii \cite{tatarskii1971effects}) models can be used to determine $C_n^2$. Alternatively, NWP software can directly solve for $C_n^2$ using the turbulent kinetic energy (TKE), see \cite{basu2020mesoscale,masciadri2017optical}. \rev{There also exist other approaches to characterize the scintillation index $\sigma_I^2$ or the variance $\sigma_\chi^2$ without relying on $C_n^2$, such as statistical models with probabilistic distributions (e.g. \cite{andrews2005,karasawa1988new}). Approaches involving wave propagation in inhomogeneous media are also possible, for example making use of phase screens (e.g. \cite{fabbro2013scintillation}), but are not considered here.} 

\rev{Regarding the determination of $C_n^2$ profiles based on meteorological quantities, focus will be given here to radiosonde measurements as they have an improved spatial resolution compared to NWP simulation data, especially in the vertical direction \cite{martini2017derivation}.}

For microwave scintillation, \cite{vasseur1999prediction} presents a probabilistic model of $C_n^2$ deduced from radiosondes that does not rely on empirical relationships derived from particular experiments. In \cite{vanhoenacker2017measurement}, a $C_n^2$ approach relying on its statistical definition is introduced to compute complementary cumulative distribution functions (CCDF) of microwave scintillation. This statistical definition approach is further extended in this work and applied to optical scintillation.

In optics, one can use radiosonde measurements to parameterize $C_n^2$ models, such as the Dewan model \cite{dewan1993model} or the Trinquet-Vernin model \cite{trinquet2007statistical}. Equivalently, analytical expressions of $C_n^2$ can be used with meteorological quantities coming from radiosonde measurements. Usually, such expressions involve the computation of vertical gradients based on the Tatarskii model of $C_n^2$ \cite{tatarskii1971effects}. Even though this approach has a theoretical basis, models of the refractive index gradient $M$ and the outer scale of turbulence $L_0$ must still be chosen, as for example in \cite{abahamid2004optical,montoya2017modeling}. This is also the approach followed in \cite{bi2020estimating,belu2012comparison}, where $C_n^2$ profiles computed with meteorological data from radiosonde measurements are then compared to $C_n^2$ measurements coming from thermosondes. A similar model is used in \cite{qing2020mesoscale} but using meteorological quantities coming from NWP simulations instead of radiosonde measurements. \rev{Alternative expressions using the Thorpes or the Ellison scales instead of the outer scale $L_0$ are also possible, as applied in \cite{gavrilov2005turbulence,basu2015simple}, and \cite{wu2020simple}.}

In this paper, $C_n^2$ is derived from its statistical definition involving the optical refractive index. This latter is computed based on meteorological quantities coming from radiosondes. This is the approach highlighted in red in Fig.~\ref{fig:schematic}, \revtwo{that is motivated by its simplicity.} Nevertheless, it provides time- and location-dependent $C_n^2$ profiles based on meteorological quantities at a given time and at a specified location. \rev{Hence, the following contributions are added in this paper:}
\begin{enumerate}
    \item An optical $C_n^2$ model relying on its statistical definition is presented, with all steps explained to extract the refractive index fluctuations based on vertical profiles of temperature and pressure only. 
    \item The presented model can be applied to any place in the world where high-density radiosonde profiles are available. It is therefore not limited to astronomical observation sites and is of particular interest for optical communication ground terminals.
    \item A fitting of parametric profiles is also conducted, enabling to obtain site-specific average $C_n^2$ models that can be useful for optical site selection.
\end{enumerate}

First, Section \ref{sec:cn2} presents the $C_n^2$ model\rev{s} used in this paper. \rev{They are then validated in Section \ref{sec:val} using radiosonde observations at Trappes (France) and Hilo, HI (USA), as well as thermosonde measurements from the T-REX campaign. \revtwo{Finally, Section \ref{sec:dis} summarizes the physical limitations of the presented model.}}

\section{Description of $C_n^2$ model}
\label{sec:cn2}
\begin{figure*}
    \centering
    \includegraphics[width=0.99\textwidth]{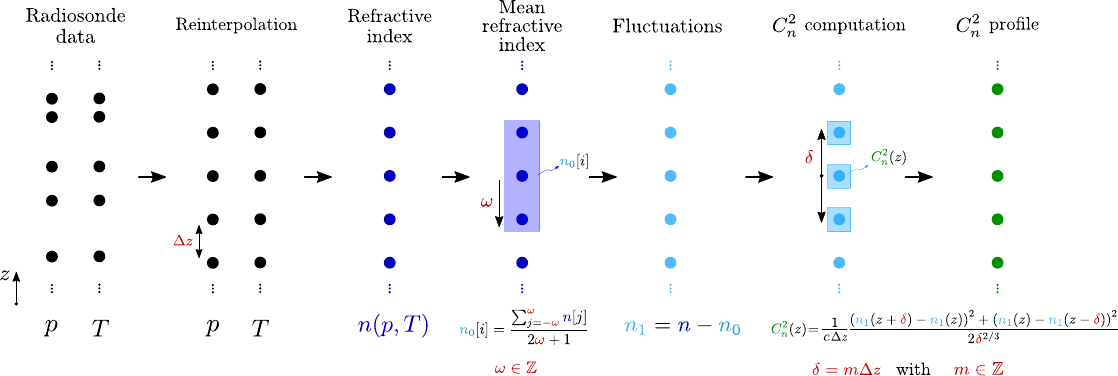}
    \caption{Followed approach to extract $C_n^2$ from its statistical definition. \textit{The three parameters ($dz$, $\omega$ and $\rho$) that can be chosen are depicted in red; $c$ is a calibration parameter.}}
    \label{fig:method}
\end{figure*}

The $C_n^2$ models presented here mainly depend on the theory that describes the refractive index as a random field. The latter is firstly reminded up to the introduction of $C_n^2$. \rev{The Tatarskii $C_n^2$ model is then presented and our alternative approach relying on the $C_n^2$ statistical definition is detailed.}

\subsection{The refrative index as a random field}
In the framework of turbulence, the refractive index $n$ is a stochastic function of time ($t$) and space ($\bm r=(r_x, r_y, r_z)$), i.e. a space-time random field, denoted $n(\bm r,t)$. Commonly, it is written \cite{andrews2005}:
\begin{align}
     n(\bm r,t) \:=\: n_0(\bm r,t) + n_1(\bm r,t),
     \label{eq:refr_index}
\end{align}
where $n_0(\bm r,t)=\langle n(\bm r,t) \rangle$ is the mean value of $n(\bm r,t)$, and $n_1(\bm r,t)$ corresponds to the fluctuation of the refractive index around its mean. By definition, $\langle n_1(\bm r,t) \rangle = 0$. \revtwo{All means are expressed as ensemble averages, e.g. \begin{align}
    n_0(\bm r,t)=\langle n(\bm{r},t)\rangle \:=\: \int_{-\infty}^{+\infty} n \: T_n(n,\bm{r},t) \:\mathrm{d}n,
\end{align} with $T_n(n,\bm{r},t)$ the probability density function of the refractive index.}

The optical refractive index can be expressed as a function of atmospheric quantities, namely the pressure and the temperature \cite{andrews2005}:%p.63, also in Wheelon 2001 p53
\begin{align}
    n \:=\: 1 + 77.6 \times 10^{-6}\left(1+7.52\times 10^{-3} \lambda^{-2}\right)\frac{p}{T}.
    \label{eq:n}
\end{align}
 In (\ref{eq:n}), $p$ is the pressure in hectopascal [hPa] and $T$ is the temperature in kelvin [K]. They can both be function of the position $\bm{r}$ and the time $t$. The wavelength $\lambda$ is in micrometer [$\mu$m] and is often considered to be 0.5 $\mu$m, leading to the following approximate expression:
 \begin{align}
     n\:\approx\: 1 + 79\times 10^{-6} \frac{p}{T}.
     \label{eq:napp}
 \end{align}

Therefore, fluctuations of pressure and temperature lead to fluctuations of the optical refractive index. Disregarding time variations of the refractive index, the structure function of the refractive index fluctuations between two positions in space ($\bm{r}_1$ and $\bm{r}_2$) is \cite{tatarskii1971effects}
\begin{align}
    D_n(\bm{r}_1,\bm{r}_2)\:=\: \left \langle \left ( n_1(\bm{r}_1)-n_1(\bm{r}_2)\right)^2\right \rangle.
\end{align}

Assuming local homogeneity, the structure function depends only on the position difference, i.e. $\bm{\rho}=\bm{r}_1-\bm{r}_2$. Adding the isotropy hypothesis, only the absolute distance difference~$\rho~=~|\bm{\rho}|$ matters and not its orientation. With these two assumptions, the structure function is now expressed as in \cite{ishimaru1978wave}:
\begin{align}
    D_n(\rho)\:=\: \left \langle \left ( n_1(\bm{r}+\bm{\rho})-n_1(\bm{r})\right)^2\right \rangle.
    \label{eq:Dth}
\end{align}

Based on the Kolmogorov cascade theory of turbulence, the structure function is shown to be proportional to $\rho^{2/3}$, the proportionality factor being the refractive index structure constant $C_n^2$ [m$^{-2/3}$]. This is only valid between the inner scale $l_0$ and the outer scale $L_0$, that depict the range where turbulent eddy energy is transferred from larger eddies to smaller eddies (inertial range) \cite{andrews2005,ishimaru1978wave,tatarskii1971effects}. Thus the structure function can be written as
\begin{align}
    D_n(\rho)\:=\:C_n^2 \:\rho^{2/3} \qquad  \text{for} \qquad l_0 < \rho < L_0.
    \label{eq:DKol}
\end{align}

Equation (\ref{eq:DKol}) is associated to the Kolmogorov power spectrum $\Phi_n(\kappa)\:=\:0.033 \:C_n^2 \: \kappa^{-11/3}$ for $1/L_0 \ll \kappa \ll 1/l_0$, where $\kappa$ is the spatial frequency. The link between these quantities is given by the following integral \cite{andrews2005}:
\begin{equation}
        D_n(\rho)\:=\:8\pi \int_0^{+\infty} \kappa^2\:\Phi_n(\kappa)\:\left(1-\frac{\sin(\kappa \rho)}{\kappa \rho}\right)\:d\kappa.
        \label{eq:Dnum}
\end{equation}

In practice, the turbulence characteristics vary with the altitude $z$ and so does $C_n^2$. This leads to a $C_n^2$ profile denoted $C_n^2(z)$.

\rev{\subsection{Tatarskii-based $C_n^2$ model}
\revtwo{Built} on Kolmogorov theory, an expression of $C_n^2$ as a function of the outer scale $L_0$ and the refractive index vertical gradient $M=\frac{\partial n}{\partial z}$ has been presented by \cite{tatarskii1971effects}:
\begin{equation}
    C_n^2 = a^2 L_0^{4/3} M^2,
    \label{eq:TatarskiCn2}
\end{equation}
where $a^2$ is usually set to 2.8 \cite{beland1988deterministic}. Equation (\ref{eq:TatarskiCn2}) is quite generic and require\revtwo{s} expressions for $L_0$ and $M$. In the following, $M$ is chosen as in \cite{cherubini2013another},
\begin{equation}
    M=-\frac{80\times 10^{-6}p}{T \theta} \frac{\partial\theta}{\partial z},
    \label{eq:M}
\end{equation}
with $\theta$ the potential temperature. For $L_0$, an extension of the \cite{dewan1993model} model, named HMNSP99, is used, as presented in \cite{ruggiero2002forecasting,wu2020simple,xu2022analysis}:
\begin{equation}
    L_0^{4/3} = 0.1^{4/3} \times 10^Y,
    \label{eq:L0}
\end{equation}
with
\begin{equation}
    Y = \begin{cases} 0.362 + 16.728 ~S - 192.347~ \frac{\partial T}{\partial z} \:\:\text{in the troposphere,} \\ 0.757 + 13.819 ~S - 57.784  ~\frac{\partial T}{\partial z} \:\:\text{in the stratosphere.}\end{cases}
\end{equation}
This is an empirical model derived from measurements in New Mexico, where $S$ is the vertical shear of the horizontal wind velocity. 
In the rest of this paper, the $C_n^2$ model from (\ref{eq:TatarskiCn2}) with (\ref{eq:M}) and (\ref{eq:L0}) will be referred to as the Tatarskii model, and serve as a literature-based model useful to validate the new approach presented in \revtwo{the} next section.}

\subsection{\rev{$C_n^2$ from statistical definition}}
\label{sec:method}
\rev{An alternative approach to obtain $C_n^2$ can make use of its statistical definition. Indeed,} in the inertial range, (\ref{eq:Dth}) and (\ref{eq:DKol}) can be used to extract $C_n^2$ from
\begin{align}
    C_n^2 \:=\:\frac{\left \langle \left ( n_1(\bm{r}+\bm{\rho})-n_1(\bm{r})\right)^2\right \rangle}{\rho^{2/3}} \qquad  \text{for} \qquad l_0 < \rho < L_0.
    \label{eq:Cn2}
\end{align}

Computation of (\ref{eq:Cn2}) is not straightforward and implies several steps detailed below and depicted in Fig.~\ref{fig:method}.

First, one assumes to have access to vertical profiles of the pressure $p$ and the temperature $T$. Such profiles are provided by radiosondes for example, as shown on the left part of Fig.~\ref{fig:method}. The vertical $z$-direction is altitude whereas black points depict positions where measurements of pressure and temperature are available. In the case of radiosondes, vertical sampling is unlikely to be equispaced. Data are then resampled to a constant vertical spacing denoted $\Delta z$ \rev{using linear interpolation. Indeed, the last step of our $C_n^2$ computation requires a fixed vertical spacing between the points, see (\ref{eq:Cn2est}) detailed below.}

Next, the refractive index is computed using (\ref{eq:n}) or (\ref{eq:napp}) depending on the knowledge of the wavelength and the choice of taking it into account \revtwo{(the variations of the optical $C_n^2$ with the wavelength can therefore be modeled with this approach)}. This provides a vertical profile of the refractive index \rev{(third step in Fig.~\ref{fig:method})}.

In (\ref{eq:Cn2}), only the fluctuations $n_1$ of the refractive index are involved but not directly the refractive index $n$ itself. Thus, there is a need to extract the refractive index mean $n_0$, and then obtain the fluctuations with $n_1=n-n_0$.
The mean should be computed locally based on neighboring points. For this reason, a window mean of size $2\omega+1$ is used, with $\omega \in \mathbb{Z}$ \rev{(steps 4 and 5 in Fig. \ref{fig:method})}.

Eventually, the $C_n^2$ profile is computed using
\begin{equation}
    C_n^2(z)\:=\:\frac{1}{c~\Delta z}\frac{\left(n_1(z+\delta)-n_1(z)\right)^2+\left(n_1(z)-n_1(z-\delta)\right)^2}{2~\delta^{2/3}}.
    \label{eq:Cn2est}
\end{equation}
The ensemble average in (\ref{eq:Cn2}) is substituted by a three-point average between points spaced by a vertical spacing $\delta$. This vertical spacing is a multiple of the vertical resolution $\Delta z$ (see Fig. \ref{fig:method}). \rev{Due to the vertical sampling, a factor $\Delta z$ arises at the denominator to ensure the homogeneity of (\ref{eq:Cn2est}). This is justified in \ref{app:A}.} Finally, $c$ is a calibration parameter named \textit{scale factor}. Its origin is motivated in Section \ref{sec:simu} and, as it is shown in the next sections, this parameter does not seem to depend on the location.

With the presented approach, three parameters must be chosen:
\begin{enumerate}
    \item the vertical spacing $\Delta z$ used for resampling the data;
    \item the integer $\omega$ depicting the window size used for the computation of the mean refractive index;
    \item the vertical spacing $\delta=m\: \Delta z$ with $m\in \mathbb{Z}$ used for computing the average in the $C_n^2$ expression.
\end{enumerate}

At this stage, some limitations of the approach can already be identified:
\begin{enumerate}
    \item The computation of the local refractive index mean $n_0$ to extract the fluctuations is difficult. Indeed, the window for computing the mean must not be too large as the refractive index mean value varies with altitude due to variations of the mean pressure and temperature with altitude.
    \item The available vertical resolution of the input meteorological data can also limit the approach and the choice of $\Delta z$. If the resolution is too low, it is unlikely that the computed mean will be representative of the local refractive index mean $n_0$. Furthermore, the choice of $\Delta z$ must be related to the vertical extent of the turbulent structures to be identified. For example, large $\Delta z$ (e.g 400 m) are useful to obtain average profiles smoothed out of noisy artefacts but do not enable to detect thin turbulent layers.
     \item The first point of the profile is ideally located at an altitude $\delta+\omega\:\Delta z$ in order to compute the mean refractive index and $C_n^2$ following Fig. \ref{fig:method}. Otherwise, neighboring points are missing to perform the computations. This means that the presented model is unable to depict close to ground turbulence, especially if $\Delta z$ is large.
    \item The presented approach is only valid between the inner scale $l_0$ and outer scale $L_0$. The inner scale is unlikely to be an issue as it is on the order of millimeter or centimeter \cite{andrews2005}. On the contrary, the outer scale spans from meters to hundreds of meters \cite{coulman1988outer,lukin2005outer} and can limit our computation (\revtwo{see Section \ref{sec:simu} and Section \ref{sec:dis}}). 
\end{enumerate}

\subsection{Synthetic data}
\label{sec:simu}
In order to validate the \rev{statistical definition} approach and determine the impacts of some parameters, such as $\omega$ or the outer length $L_0$, a preliminary analysis is performed with synthetic data. \rev{A study of the limitations related to the outer scale $L_0$ is of particular interest since, in practical scenarios, this scale is often found to be smaller than 10 meters \cite{beland1988deterministic,coulman1988outer,abahamid2004optical}. However, in Section \ref{sec:val}, radiosonde observations with a vertical resolution on the order of 10 meters will be used, with a resampling distance $\Delta z$ between 25 and 400 meters. Hence, the assumption about the inertial range, required for applying (\ref{eq:Cn2}), may not hold, and this impact is studied here.}

\rev{For this study with synthetic data,} one-dimensional (1D) fluctuations of the refractive index have been generated through white noise coloring to obtain the desired properties (i.e.~the desired spatial power spectrum) on the resulting fluctuations. \rev{In order to include the limitations imposed by the outer scale $L_0$ in the simulation, the von K\'arm\'an spectrum has been chosen \cite{andrews2005}:
  \begin{equation}
    \Phi_{n}(\kappa)\:=\:\frac{0.033 ~C_n^2}{(\kappa^2+\kappa_0^2)^{11/6}},
\end{equation}
with $\kappa_0=1/L_0$. The associated 1D power spectrum $V_n(\kappa)$ is found thanks to \cite{andrews2005}:
\begin{equation}
    \Phi_{n}(\kappa)\:=\:-\frac{1}{2\pi \kappa}\frac{dV_n(\kappa)}{d\kappa}.
\end{equation}
}

\begin{figure}
    \centering
    \includegraphics[width=0.5\textwidth]{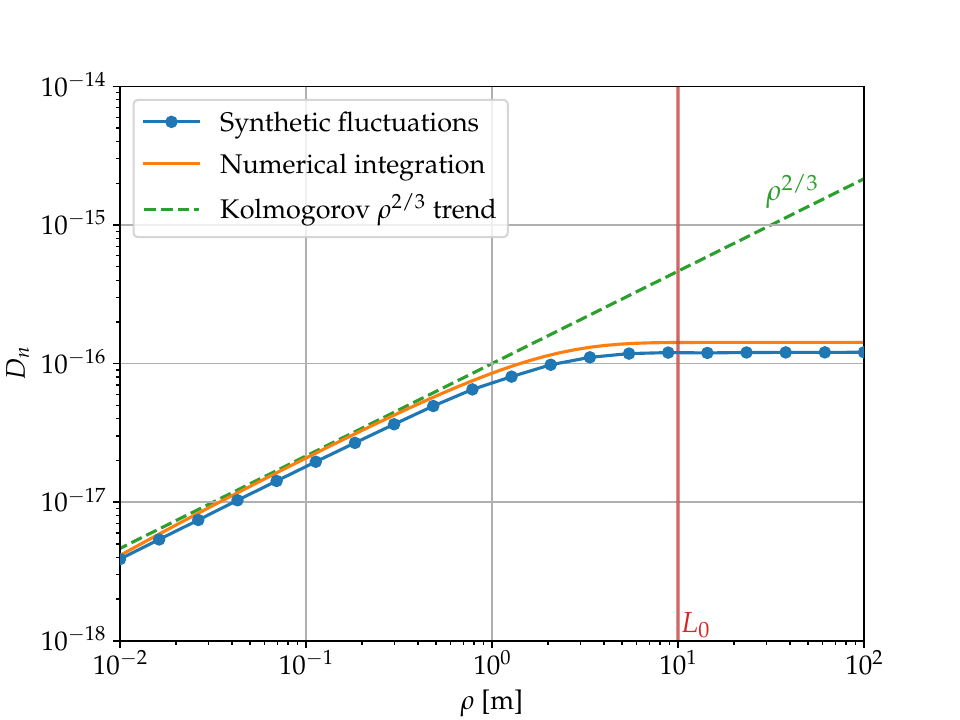}
    \caption{Structure function from 1D simulations.}
    \label{fig:struct_funct}
\end{figure}

Figure \ref{fig:struct_funct} represents the structure function computed from the simulated fluctuations using (\ref{eq:Dth}), substituting the ensemble average by a spatial average. The chosen parameters for the simulation are: \rev{$L_0=10$ m and $C_n^2=10^{-16}$ m$^{-2/3}$}. The structure function is computed for several spacing $\rho$ and compared to the Kolmogorov trend, from (\ref{eq:DKol}). Agreement is indeed reached in the inertial range (i.e.~between $l_0$\rev{, assumed to be $0$ here,} and $L_0$), even though the structure function tends to be slightly underestimated. Numerical integration of (\ref{eq:Dnum}) with the von K\'arm\'an spectrum is also depicted, highlighting the saturation of the structure function above the outer scale~$L_0$.

\begin{figure}
    \centering
    \includegraphics[width=0.5\textwidth]{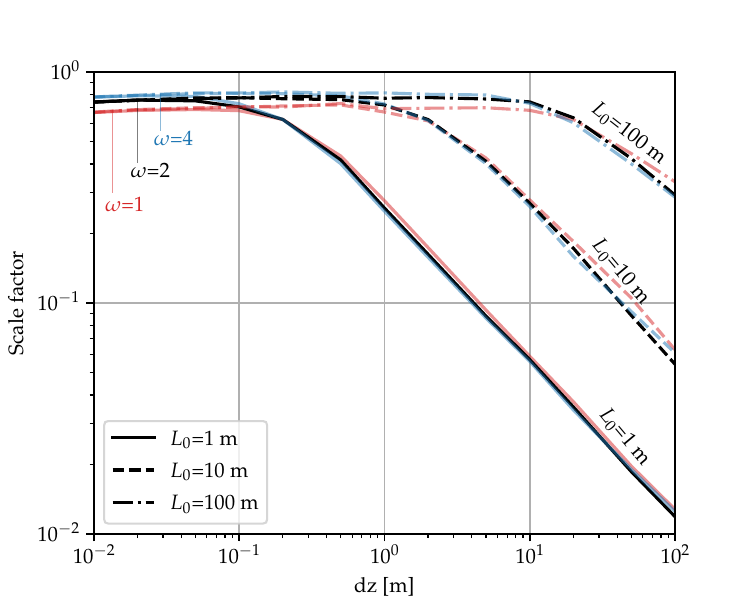}
    \caption{Scale factor for $C_n^2$ correction on synthetic data. \textit{Colors depict the choice of the window size parameter $\omega$.}}
    \label{fig:th_study}
\end{figure}

Figure \ref{fig:th_study} shows the scale factor $c$ for $C_n^2=10^{-16}$ m$^{-2/3}$ and an outer scale $L_0$ varying from 1 to 100 meters. This factor is defined as the ratio between the estimated $C_n^2$ \rev{(i.e.~the one recovered using the statistical definition method on the simulated data)} and the real $C_n^2$ (i.e.~the one applied as an input in the generation of synthetic data). $C_n^2$ is estimated assuming different spatial resolutions $\Delta z$, with $\delta=\Delta z$ and varying window size $\omega$ for the mean estimation.

For resolution sufficiently smaller than the outer scale (roughly one order \revtwo{of} magnitude), the estimated $C_n^2$ and the simulation value are close since the factor is a bit lower than one. In other words, there is a small underestimation of $C_n^2$, mostly coming from an underestimation of the fluctuations $n_1$ when extracting the mean. Indeed, neighboring fluctuations are correlated, meaning that the extracted mean does not only contain $n_0$ but also a non-zero mean of neighbored fluctuations $n_1$. Therefore, fluctuations tend to be underestimated and so is $C_n^2$. Increasing the window size $\omega$ enables to look at more points that are also further from each other, with less correlated fluctuations, improving the extracted mean $n_0$ and therefore reducing the underestimation of fluctuations.

For spacing $\Delta z$ close to or larger than $L_0$, the factor starts to decrease following a minus two-third power trend. Indeed, the validity of (\ref{eq:Cn2}) is not ensured anymore: as observed in Fig.~\ref{fig:struct_funct}, the structure function tends to saturate. Since this structure function is exactly the numerator in (\ref{eq:Cn2}), only the influence of its denominator remains, that corresponds exactly to a minus two-third power trend. %Hence, the $C_n^2$ estimation must be corrected by a factor $c$.
\rev{If observed with real-world measurements, such a decrease of the scale factor $c$ would be a hint that the outer scale $L_0$ has been reached. A possible approach to obtain the factor $c$ for $C_n^2$ profiles based on radiosonde measurements is presented in Section \ref{sec:calib}.}

\section{\rev{Validation of the $C_n^2$ models through data analysis}}
\label{sec:val}
\rev{In this section, both $C_n^2$ models presented in Section \ref{sec:cn2}, i.e.~the Tatarskii model and the statistical definition model, are applied to radiosonde measurements for validation. As a first step, the statistical definition model is calibrated thanks to the Hufnagel-Valley model.}
\subsection{\rev{Calibration using Hufnagel-Valley model}}
\label{sec:calib}
\rev{Calibration of the statistical definition model refers to the determination of the scale factor $c$ in practical scenarios.} This can be achieved by comparing obtained $C_n^2$ profiles with empirical models from the literature\rev{, such as the Hufnagel-Valley model. A generalisation of this model is presented in \cite{hardy1998adaptive}:
\begin{align}
    C_n^2(z)\:&=\:A\exp{\left(-\frac{z}{H_A}\right)}+B\exp{\left(-\frac{z}{H_B}\right)} \nonumber\\
    &+ C \left(\frac{z}{10^5}\right)^{10}\exp{\left(-\frac{z}{H_C}\right)}+ D\exp{\left(-\frac{\left(z-H_D\right)^2}{2d^2}\right)},
    \label{eq:HVgen}
\end{align}
with $z$ the altitude above sea level [m]. Equation (\ref{eq:HVgen}) involves several coefficients: $A$, $B$, $C$, and $D$, that have the units of a $C_n^2$ [m$^{-2/3}$], and $H_A$, $H_B$, $H_C$, and $H_D$, that are expressed in [m]. The first term in (\ref{eq:HVgen}) depicts the surface turbulence, the second term represents the troposphere turbulence whereas the third term defines the turbulence at the tropopause. Moreover, there is a fourth term representing isolated layers of turbulence with a Gaussian shape of thickness $d$. This term is repeated if there are several turbulent layers.}

\rev{Usually, the Hufnagel-Valley 5/7 profile (HV-5/7) is used in astronomy and for optical communications \cite{andrews2005}. Its expression is a particular case of (\ref{eq:HVgen}):
\begin{align}
    C_n^2(z)\:&=A \exp{\left(-\frac{z}{100}\right)} + 2.7\times10^{-16}\exp{\left(-\frac{z}{1500}\right)}  \nonumber \\
    &+  0.00594\left(\frac{w}{27}\right)^2\left(\frac{z}{10^5}\right)^{10}\exp{\left(-\frac{z}{1000}\right)}, 
    \label{eq:HV}
\end{align}
where $A$ is the $C_n^2$ value near the ground [m$^{-2/3}$], and $w$ is the root-mean-square high altitude wind speed [m/s]. The HV-5/7 profile is obtained with $w=21$ m/s and $A=1.7\times10^{-14}$ m$^{-2/3}$.}

\begin{figure}
    \centering
    \includegraphics[width=0.5\textwidth]{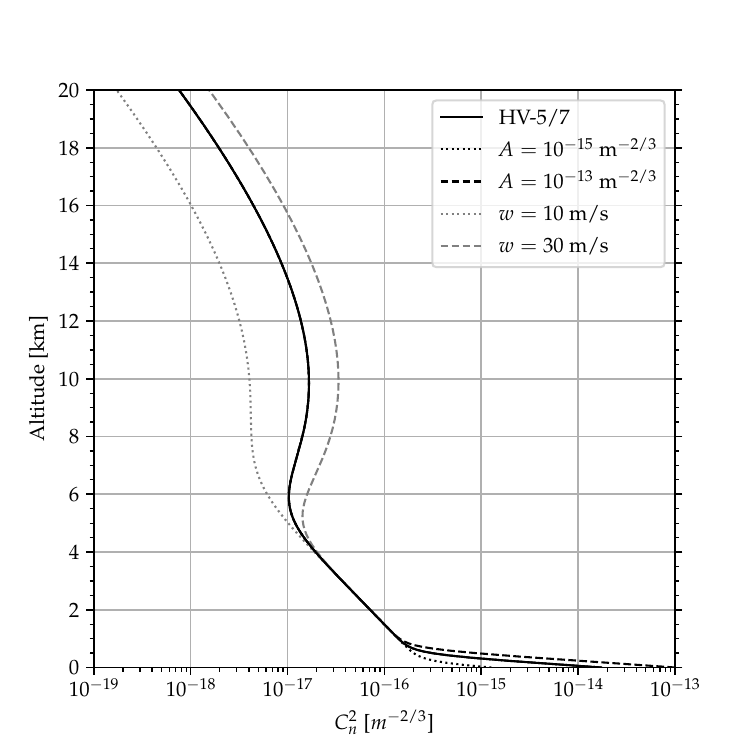}
    \caption{Hufnagel-Valley model variations with $w$ and $A$ parameters. \textit{If not specified, parameters used are the ones of HV-5/7.}}
    \label{fig:HVparam}
\end{figure}

\rev{The important observation in (\ref{eq:HV}) is the $B$ coefficient that is set to a constant value of $2.7\times10^{-16}$. It leads to a fixed range in the Hufnagel-Valley model (between 1 and 4 km of altitude), that is independent of the other parameters $w$ and $A$, as depicted in Fig. \ref{fig:HVparam}. This remarkable feature is most probably related to the relatively constant temperature and pressure mean lapse rate in the troposphere.}

\rev{When applying the statistical definition $C_n^2$ model to radiosonde data, the same constant slope (on a semilogarithmic representation) has been observed in the 1 to 4 km region. Hence, this characteristic can be exploited to calibrate the obtained $C_n^2$ profiles in order to ensure a good fit between these profiles and the HV-5/7 profile in this region. This is illustrated at Trappes (France) and Hilo, HI (USA), in Sections \ref{sec:Trappes} and \ref{sec:Hawaii}.}

\subsection{Application to radiosonde measurements at Trappes, France}
\label{sec:Trappes}
High-density profiles from the University of Wyoming (UWYO) Atmospheric Science Radiosonde Archive\footnote{\url{http://weather.uwyo.edu/upperair/bufrraob.shtml}} have been used as input meteorological data. Data collected above Trappes (latitude 48.77° N, longitude 2.01°E, altitude 168 meters) in France during the year 2020 have been analyzed. Radiosondes are launched twice a day, at 00h00 and 12h00 UTC. They offer a vertical resolution of 10 meters (standard deviation: 3 m) nearly independent of the altitude, and provide altitude, pressure, temperature, relative humidity, wind speed and wind direction profiles.
Moreover, radiosonde drifting due to wind has been neglected, meaning that $C_n^2$ profiles obtained are considered at zenith while in fact they are obtained along the radiosonde trajectory. \textcolor{black}{This drifting is spatially limited since, for example, for the year 2020 at Trappes, the average drift (computed at the altitude of 20 km following \cite{laroche2013impact} with a mean ascent rate of 6 m/s) is close to 50 km, with a standard deviation of 28 km. It is also not expected to influence the vertical resolution as the radiosonde directly records the altitude above ground and not the distance traveled by the balloon.}

\begin{figure*}
    \centering
    \includegraphics[width=0.8\textwidth]{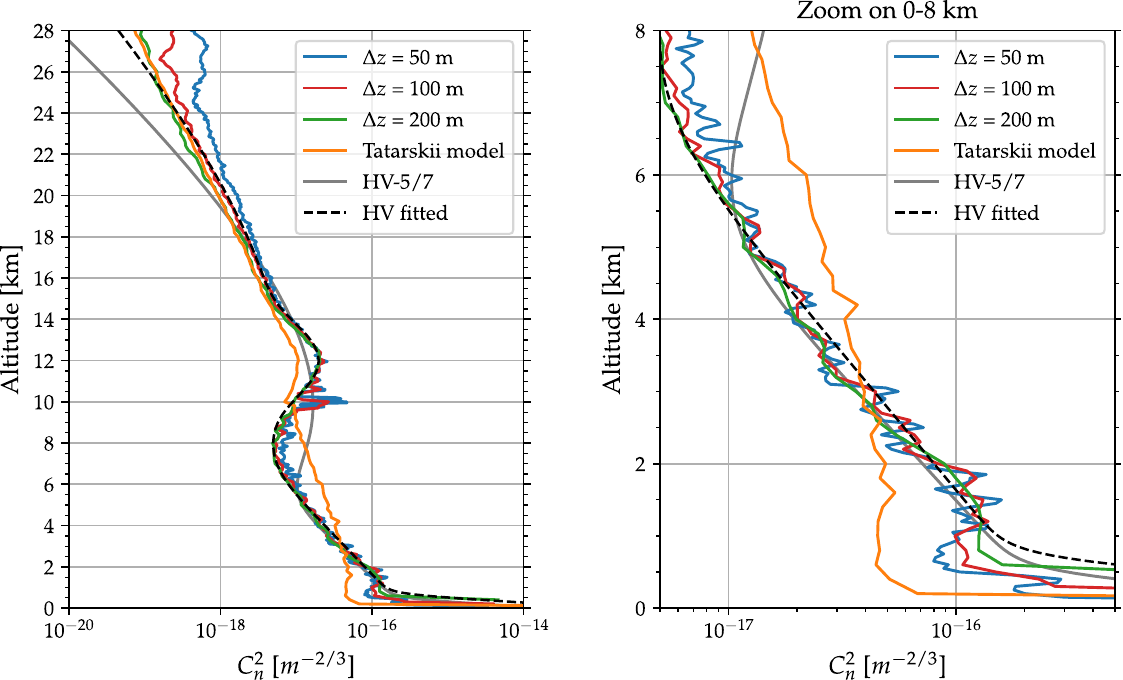}
    \caption{\rev{Average $C_n^2$ profiles at Trappes for the year 2020, after calibration.}}
    \label{fig:Trappes_average_corr}
\end{figure*}

Following the approach presented in Section \ref{sec:method}, $C_n^2$ profiles have been computed for each available measurement, neglecting the days where the radiosonde did not reach the altitude of 30 km. In total, 630 different $C_n^2$ profiles have been obtained and computed with different vertical spacings $\Delta z$. The averages of these profiles are presented in Fig. \ref{fig:Trappes_average_corr} for $\Delta z$ of 50, 100 and 200 meters. For all $C_n^2$ computations, parameter $\omega$ has been set to 2 whereas $\delta$ was equal to the $\Delta z$-spacing. The approximate model of $n$ from (\ref{eq:napp}) has been used. The altitude corresponds to the altitude above ground (and not above the mean sea level). \rev{For comparison, the Tatarskii $C_n^2$ model has also been added, directly computed from the radiosonde measurements and using a troposphere maximum altitude of 10 km in (\ref{eq:L0}).}

All three profiles of Fig. \ref{fig:Trappes_average_corr} coming from the $C_n^2$ statistical definition have been calibrated with their associated scale factor. Computation of these calibration ratios has been performed using the methodology explained in Section \ref{sec:calib}, i.e. by ensuring agreement with the HV-5/7 profile in the 1-4 km region. Such agreement is indeed achieved, \rev{as can be seen on the zoom in Fig.~\ref{fig:Trappes_average_corr}}, using the factors given in Table \ref{tab:scale_factor}, with $\omega=2$ and $\delta=\Delta z$. They have been obtained by first computing the ratio between the modeled $C_n^2$ value and the HV-5/7 $C_n^2$ value for each point in the 1-4 km region, then taking the average. The standard deviation is also given.

These scale factors are independent of the altitude and the initial radiosonde vertical resolution. They vary slightly with the window size parameter $\omega$ and the vertical spacing $\delta$. \rev{Surprisingly, the values obtained do not show the $\Delta z^{-2/3}$ trend that is observed \revtwo{in} the simulation values in Fig. \ref{fig:th_study} for $\Delta z$ larger than the outer scale $L_0$. This may hint that the outer scale has not been reached yet, and that constraints on $\rho$ in (\ref{eq:DKol}) can be relaxed, or, at least, are partially hidden by the calibration with the scale factor $c$.} 
In the following, the scale factor $c=0.5$ is chosen for $\omega=2$ and $\delta=\Delta z$, for all $\Delta z$. 

\begin{table}
    \centering
    \caption{Scale factors depending on $\Delta z$ - mean value and standard deviation.}
\begin{tabular}{l c c c}
\hline
\multicolumn{1}{l}{\multirow{2}{*}{$\Delta z$ {[}m{]}}} & \multicolumn{3}{c}{Scale factors $c$}                                                                                   \\ %\cline{2-4} 
\multicolumn{1}{l}{}                            & $\omega=1$, $\delta=\Delta z$ & $\omega=2$, $\delta=\Delta z$ & $\omega=2$, $\delta=2 \Delta z$ \\ \hline
25  & 0.47~\footnotesize{$\pm0.12$} & 0.60~\footnotesize{$\pm0.12$} & 0.60~\footnotesize{$\pm0.1$}  \\
50  & 0.37~\footnotesize{$\pm0.06$} & 0.53~\footnotesize{$\pm0.08$} & 0.57~\footnotesize{$\pm0.08$}  \\ 
100  & 0.36~\footnotesize{$\pm0.04$} & 0.53~\footnotesize{$\pm0.05$} & 0.58~\footnotesize{$\pm0.06$}  \\
200  & 0.36~\footnotesize{$\pm0.03$} & 0.51~\footnotesize{$\pm0.04$} & 0.56~\footnotesize{$\pm0.05$}  \\ 
400  & 0.35~\footnotesize{$\pm0.01$} & 0.48~\footnotesize{$\pm0.01$} & 0.51~\footnotesize{$\pm0.02$}  \\ 
\hline
\end{tabular}
    \label{tab:scale_factor}
\end{table}

With the calibration, \rev{$C_n^2$ profiles coming from the statistical definition} in Fig. \ref{fig:Trappes_average_corr} tend to superimpose. They all contain a large ground $C_n^2$ value and a fixed slope from 1 to 6 kilometers of altitude, \rev{that exactly correspond\revtwo{s} to the same slope as in the HV-5/7 model. The slope in this region for the Tatarskii model is slightly different}. There is then a bump in the $C_n^2$ profile, around 12 km of altitude, that is related to the altitude of the tropopause and the associated constant temperature and large wind shears. \rev{This bump is also observed in the Tatarskii model.} After the tropopause, $C_n^2$ slowly decreases again and starts experiencing some fluctuations related to the reinterpolation process. This mostly comes from the small variations of the atmospheric quantities at these altitudes that cannot be recorded by the radiosondes as well as from the uneven and larger vertical sampling distance of radiosondes at high altitudes. 

Interestingly, the shape of $C_n^2$ profiles is not similar to the HV-5/7 model above the altitude of 6 kilometers. This is because the free atmosphere part of the Hufnagel-Valley model has been derived based on radiosonde measurements over New Mexico \cite{HVchapter6} and is therefore sensitive to the location where the measurements have been made.

Fitting the model in (\ref{eq:HVgen}) leads to the dashed line in Fig. \ref{fig:Trappes_average_corr}, denoted as \textit{HV fitted}, with the coefficients presented in Table \ref{tab:coeff}. The fitting has been performed by minimizing the least square error with the statistical definition $C_n^2$ profile obtained from radiosonde measurements with $\Delta z$=200 m. The four parameters in blue in Table \ref{tab:coeff} have been set prior to the least square fitting. As seen in Fig. \ref{fig:Trappes_average_corr}, good parameterization of the generalized HV profile enables to analytically model the mean $C_n^2$ profile at a particular location (Trappes, in this case). This offers very simple expressions of average $C_n^2$ profiles that can be obtained at different locations and compared, e.g. to perform site selection for building optical communication stations.

\begin{table}
    \centering
    \caption{Parameters for generalized HV model in (\ref{eq:HVgen}).}
    \begin{tabular}{l c c}
    \hline
         Parameters & Values for Trappes & Values for Hilo \\
         \hline
         $A$ [m$^{-2/3}$] & $1.32\times 10^{-13}$ & $4.66\times 10^{-14}$ \\
         
         \textcolor{blue}{$B$ [m$^{-2/3}$]} & \textcolor{blue}{$2.7\times 10^{-16}$} & \textcolor{blue}{$2.7\times 10^{-16}$} \\
         
         $C$ [m$^{-2/3}$] &  $2.07\times 10^{-4}$ &  $2.96\times 10^{-5}$\\
         
         $D$ [m$^{-2/3}$] &  $1.37\times 10^{-17}$ & $4.67\times 10^{-18}$ \\
         
         \textcolor{blue}{$H_A$ [m]} & \textcolor{blue}{$100$} & \textcolor{blue}{$100$}\\
         
         $H_B$ [m] &  $1645$ &  $2006$\\
         
         $H_C$ [m] &  $1200$ & $1340$\\
         
         \textcolor{blue}{$H_D$ [m]} & \textcolor{blue}{$12000$} & \textcolor{blue}{$17000$}\\
         
         \textcolor{blue}{$d$ [m]} & \textcolor{blue}{$1200$}  & \textcolor{blue}{$1700$}\\
         
         $E$ [m$^{-2/3}$]  &  / & $1.59\times 10^{-16}$\\
        
         \textcolor{blue}{$H_E$ [m]} & \textcolor{blue}{/} & \textcolor{blue}{$2200$}\\
         
         \textcolor{blue}{$e$ [m]} & \textcolor{blue}{/}  & \textcolor{blue}{$300$}\\
         \hline
    \end{tabular}
    \label{tab:coeff}
\end{table}

\subsection{Application to radiosonde measurements at Hilo, HI (USA)}
\label{sec:Hawaii}
Generalization of the \rev{statistical definition} $C_n^2$ model presented in Section \ref{sec:method} and applied to Trappes in Section \ref{sec:Trappes} is further studied by using radiosonde measurements at other locations. In this section, the radio sounding station of Hilo (latitude 19.717° N, longitude 155.05°W, altitude 10 meters) at Hawaii (USA), has been chosen, namely because of its completely different location with respect to Trappes and because of the availability of high-density profiles on the University of Wyoming website. A total of 677 radiosonde profiles for the year 2020 have been exploited, offering a vertical resolution of 5 meters (standard deviation: 1 m). 

\begin{figure*}
    \centering
    \includegraphics[width=0.8\textwidth]{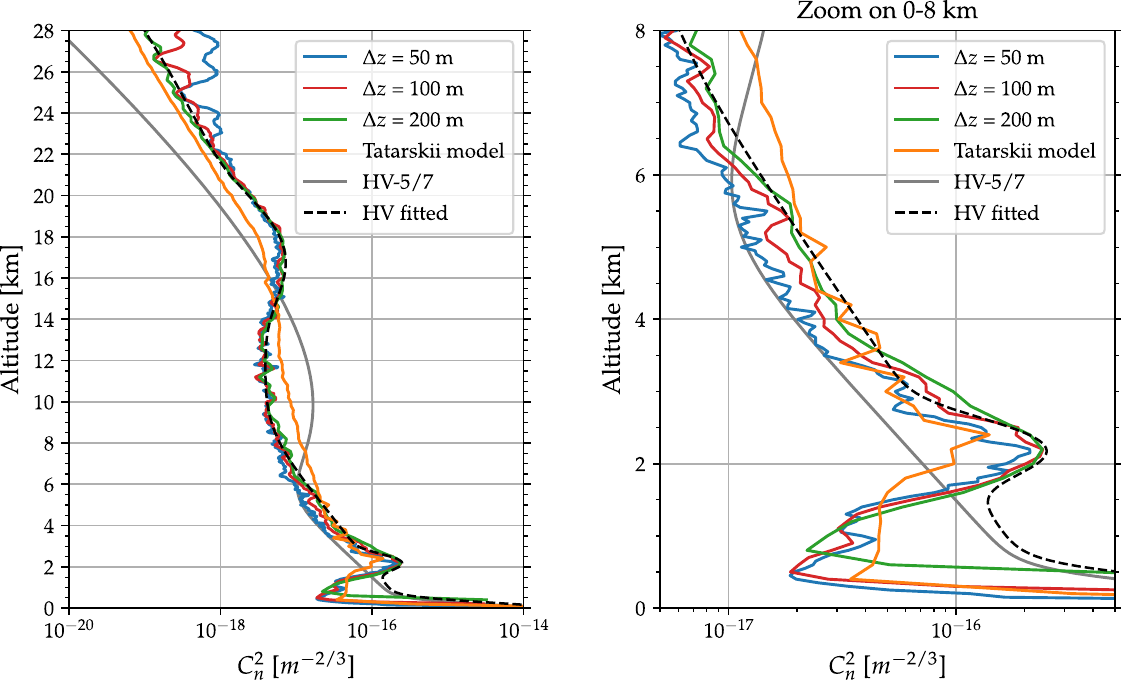}
    \caption{\rev{Average $C_n^2$ profiles at Hawaii for the year 2020, after calibration.}}
    \label{fig:Hawaii_average}
\end{figure*}

The mean $C_n^2$ profiles computed with the same $\omega$, $\delta$ and scale factors as for Trappes (see Table \ref{tab:scale_factor}) are presented in Fig. \ref{fig:Hawaii_average}. Good agreement is achieved between the profiles at different $\Delta z$ after applying the Trappes scale factors, hinting at the location independence of the factors. \rev{Indeed, disregarding the isolated turbulent layer at 2-km of altitude, \revtwo{which} origin is discussed below, the slope of the statistical definition $C_n^2$ profiles is the same as the slope of the Hufnagel-Valley profile ($B=2.7\times 10^{-16}$), and quantitative agreement with HV-$5/7$ is nearly reached using Trappes scaling factors. Furthermore, in Fig. \ref{fig:Hawaii_average},} the generalized HV model in (\ref{eq:HVgen}) has also been fitted, adding another term to model an isolated turbulent layer. All fitted parameters are given in Table \ref{tab:coeff}. Compared to the mean profiles at Trappes from Fig. \ref{fig:Trappes_average_corr}, there are two striking differences:
\begin{enumerate}
    \item The tropopause and its associated turbulent layer is located at a higher altitude, close to 17 km, as given by the parameter $H_D$. The difference with the classical HV-5/7 model is substantial.
    \item In addition to the tropopause Gaussian turbulent layer, there is a second isolated layer close to the ground (2 km altitude), modeled by the parameters $E$ (magnitude), $H_E$ (altitude of middle point) and $e$ (thickness). Even though the generalized HV model can depict this turbulent layer, it is unable to represent the reduced turbulence strength below the layer. \rev{This layer is also identified in the Tatarskii model, that has been applied using a troposphere's limit of 16 km in (\ref{eq:L0}).}
\end{enumerate}

\begin{figure}
    \centering
    \includegraphics[width=0.5\textwidth]{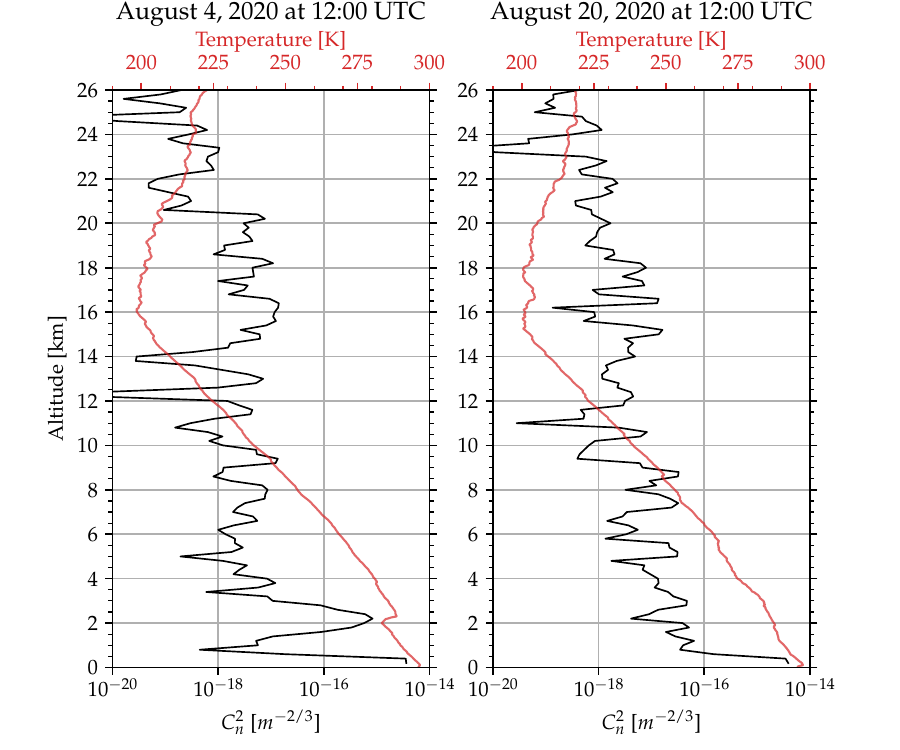}
    \caption{Two examples of day $C_n^2$ profiles at Hilo, Hawaii. \textit{The $C_n^2$ model used is the statistical definition model with $\Delta z$ = 200 m.}}
    \label{fig:Hawaii_2_profiles}
\end{figure}

The origins of this second turbulent layer can be seen in Fig. \ref{fig:Hawaii_2_profiles}. $C_n^2$ profiles computed with \rev{the statistical definition and} a $\Delta z$ of 200 m are presented at Hilo for two different days. On the left graph, a turbulent layer at an altitude of 2 km is noticeable, with $C_n^2$ reaching nearly $10^{-15}$ m$^{-2/3}$. This layer is associated with a tropospheric temperature inversion arising at this altitude. Indeed, the temperature, depicted in red, starts to increase before decreasing again. Sharp variations of temperature will influence $C_n^2$ obtained from its statistical definition since the temperature is involved in the computation of the refractive index. On the contrary, on the right graph of Fig.  \ref{fig:Hawaii_2_profiles}, there is no temperature inversion and no turbulent layer is identified. Links between regions of high $C_n^2$ and temperature inversions have also been identified in \cite{coulman1986observation}. Since temperature inversions close to ground are quite common at Hilo, this explains why this turbulent layer remains noticeable in the mean profile.

\subsection{\rev{Application to T-REX thermosonde measurements at Three Rivers, CA (USA)}}
\label{sec:thermosonde}

Access to thermosonde measurements from the T-REX campaign of the Air Force Research Laboratory (AFRL) has been granted \cite{TREX}. These data are of particular interest since they include radiosonde measurements as well as thermosonde measurements providing the $C_n^2$ profile. A comparison between \rev{the $C_n^2$ models} (based on radiosonde measurements) and the $C_n^2$ profile retrieved from the thermosondes is therefore possible.

The T-REX campaign took place from March 20, 2006 to April 6, 2006 in Sierra Nevada Moutains near Three Rivers, CA (launch site: latitude 36.4872° N, longitude 118.84048°W, altitude 503 meters) \cite{jumper}. The purpose of the experiments was to study higher altitude mountain waves and the associated turbulence. Radiosondes offer a vertical resolution of 10 meters (standard deviation: $3.7$ m) recording the altitude, pressure, temperature, relative humidity, wind speed and wind direction.

Thermosondes are able to compute the temperature structure function $D_T(\rho)$ directly based on its definition (similar to (\ref{eq:Dth}) but for the temperature) thanks to measurements coming from two temperature sensors spaced by a given distance $\rho$ ($\rho=1$ m for the T-REX thermosondes) and using a temporal average. Then, the temperature structure parameter $C_T^2$ is extracted, similarly to (\ref{eq:DKol}). Using the pressure and temperature from the radiosonde observations, the temperature structure parameter is finally converted to the refractive index structure parameter $C_n^2$ thanks to \cite{beland}:
\begin{equation}
    C_n^2 \:=\: \left(\frac{\partial n}{\partial T}\right)^2\:C_T^2\:=\: \left(79\times 10^{-6} \frac{p}{T^2}\right)^2\: C_T^2,
\end{equation}
with $p$ in hectopascal, $T$ in kelvin and the refractive index expression coming from (\ref{eq:napp}).

\begin{figure*}
    \centering
    \includegraphics[width=0.8\textwidth]{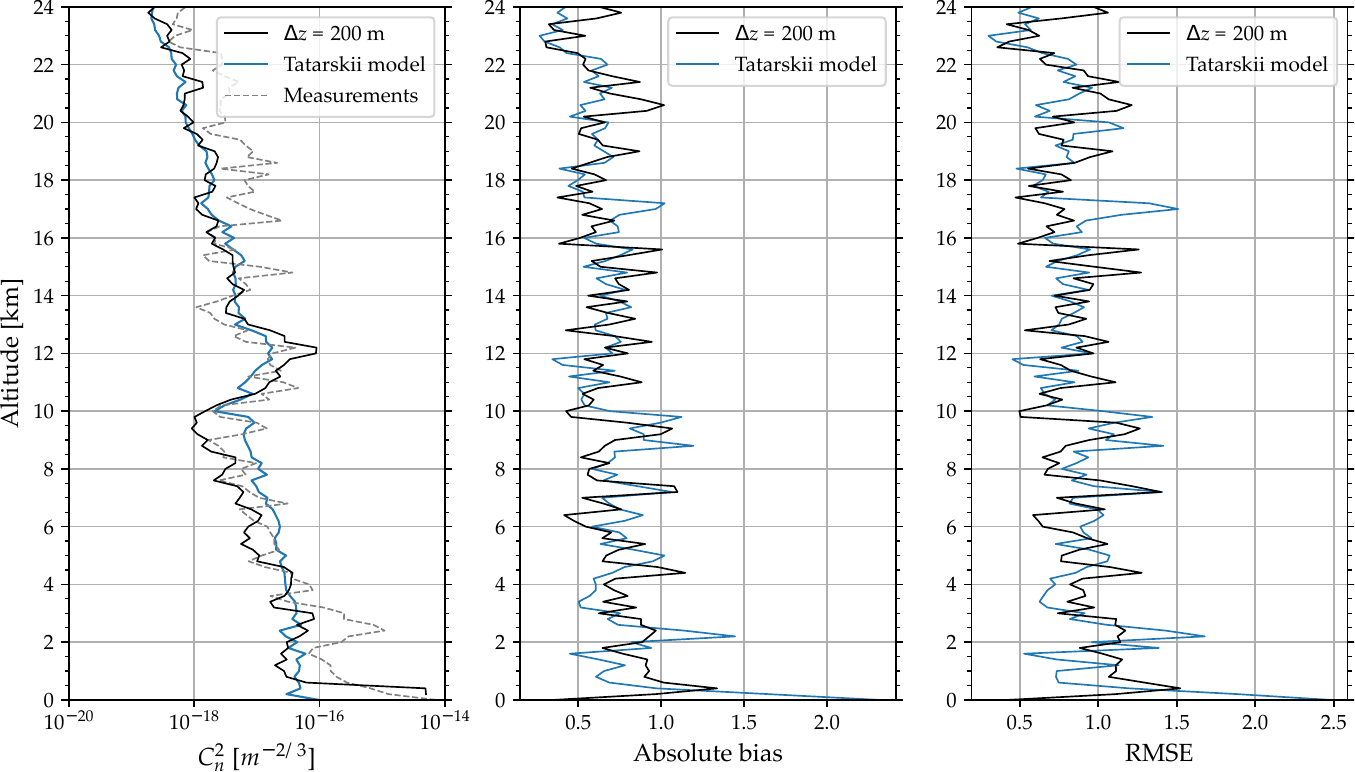}
    \caption{\rev{Average (March 20 - April 26, 2006) profiles for T-REX campaign. \textit{Left: $C_n^2$ profiles. Center: absolute bias. Right: root-mean-square error. Solid lines are the modeled $C_n^2$ based on radiosonde measurements while the dashed line corresponds to thermosonde measurements.}}}
    \label{fig:TREX_average}
\end{figure*}

\begin{figure}
    \centering
    \includegraphics[width=0.5\textwidth]{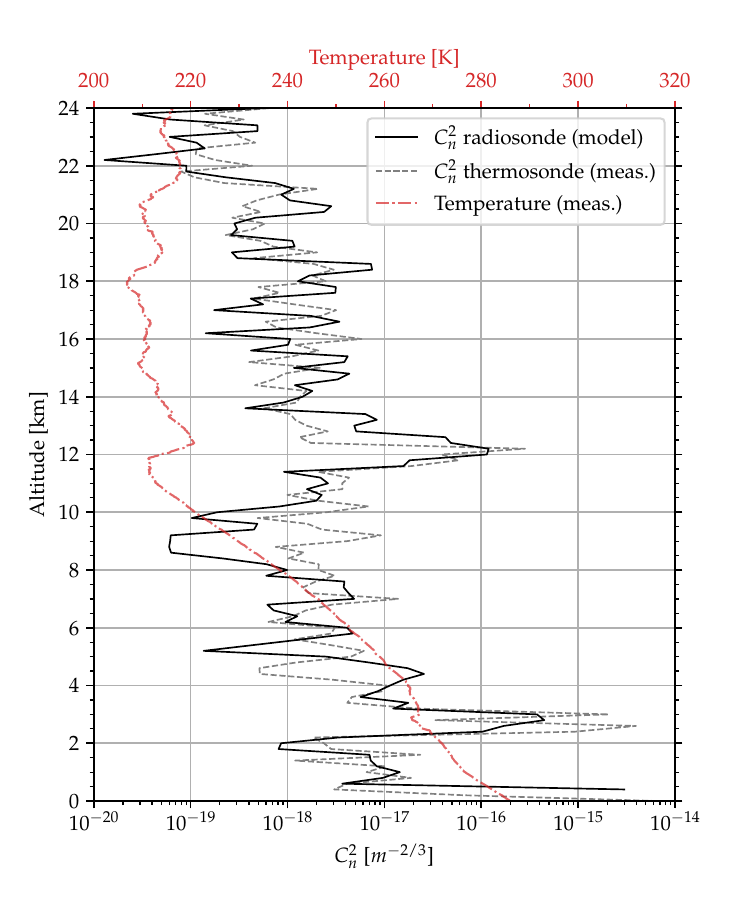}
    \caption{$C_n^2$ profiles on March 31, 2006 at 00h48 UTC (T-REX019). \textit{The $C_n^2$ model used is the statistical definition model with $\Delta z$ = 200 m.}}
    \label{fig:19}
\end{figure}

In total, measurements of 15 different flights have been exploited. \rev{Two} flight\rev{s} (\rev{T-REX011 and} T-REX039 ) ha\rev{ve} been discarded as \rev{they} showed strong turbulence up to two orders of magnitude larger than other flights. The remaining \rev{13} flights have been used to compute average $C_n^2$ profiles. 
\rev{Both $C_n^2$ models have been applied to the radiosonde measurements, with the statistical definition model using the same parameters as for Trappes (and Hilo), and a $\Delta z$ of 200 meters. The troposphere's limit is set to 10 km in the Tatarskii model.}
Regarding the thermosonde measurements, they have been binned and averaged over the same distance $\Delta z$. The average profiles can be seen in Fig. \ref{fig:TREX_average}, where the mentioned altitude is the altitude above the ground.

\rev{In the center and the right parts of this figure, two metrics have been added in order to quantify the performance of both $C_n^2$ models: the absolute bias and the root-mean-square error (RMSE). They are applied to the $\log_{10}(C_n^2)$, and their definitions are inspired from \cite{xu2022analysis}:
\begin{align}
    \text{Absolute bias}\:&=\:\sum_{i=1}^N \frac{|Y_i-X_i|}{N},\\
    \text{RMSE}\:&=\:\sqrt{\sum_{i=1}^N \frac{(Y_i-X_i)^2}{N}},
\end{align}
where $N$ is the number of measurements ($N=13$), $X_i$ represent\revtwo{s} the $i$-th measurement of $\log_{10}(C_n^2)$ and $Y_i$ is the logarithm in base 10 of the $i$-th modeled $C_n^2$ profile (either using the Tatarskii model of the statistical definition model). These metrics are function of the altitude $z$. Values obtained for both models are very similar, showing that the statistical definition $C_n^2$ model delivers $C_n^2$ estimates similar to the ones of Tatarskii-based models. Averaging over all altitudes, the absolute bias (resp. RMSE) for the Tatarskii model is 0.674 (resp. 0.842), whereas it is 0.683 (resp. 0.852) for the statistical definition model. Hence, according to these metrics, models have similar capabilities.}

\rev{Furthermore, in Fig. \ref{fig:TREX_average},} both modeled and measured $C_n^2$ profiles show the influence of the tropopause layer (close to 12 km of altitude) where they agree quite well. The decrease of $C_n^2$ in the troposphere is also identified. However, measurements show the presence of a turbulent layer close to the ground (2-3 km of altitude) that is not well depicted in the $C_n^2$ model\rev{s}. This effect comes from days \revtwo{when} the magnitude of the turbulent layer is underestimated. \rev{Indeed,} as an example, the $C_n^2$ profile recorded on March 31, 2006 is given in Fig. \ref{fig:19}%, computed with a $\Delta z=200$ m
. Isolated optical turbulent layers corresponding to temperature inversions can clearly be identified, in the thermosonde measurements as well as in the radiosonde model. This is similar to what has been observed in Fig. \ref{fig:Hawaii_2_profiles} for Hilo, HI, and in \cite{mahalov2010characterization}.

\revtwo{\section{Discussion and generalization}
\label{sec:dis}
This $C_n^2$ model, inspired by the statistical definition of $C_n^2$ in the inertial range, does not claim to be an accurate description of the physics of optical turbulence. Indeed, as previously introduced in Section \ref{sec:method}, the hypothesis of the inertial range, necessary for (\ref{eq:Cn2}) to hold, may not be fulfilled given the available resolution of radiosonde measurements. Moreover, the model involves a fitted parameter that does not have a physical justification: the scaling factor.}

\revtwo{Nevertheless, this model offers a simplified description of $C_n^2$, easy to apply for practical scenarios. Despite its simplicity, it has also been shown in Section \ref{sec:thermosonde} that the accuracy of the model is comparable to other literature-based models, as long as it is calibrated accordingly.}

\revtwo{Generalization of this approach to rely on physical scaling laws in the buoyancy range, i.e.~the range made of scales larger than the outer scale $L_0$, can be considered, and is of particular interest to obtain a model that may be free of fitting parameters. In the buoyancy range, different scaling laws of the temperature structure function have been proposed \cite{bolgiano1962structure,cot1989spectral,nastrom1997vertical}. The hypotheses behind those different scaling laws are presented in \cite{basu2022revisiting}, leading to temperature structure functions proportional to $\rho^{2/5}$ or $\rho^2$ in the buoyancy range.}

\section{Conclusion}
A simple approach to estimate the optical $C_n^2$ profile based on its statistical definition has been presented, \rev{and compared with a Tatarskii-based model. This simple approach is} made possible thanks to the use of high-density radiosonde profiles\rev{, and} relies on the optical refractive index expression, involving only the pressure and the temperature profiles. From the refractive index vertical profile, refractive index fluctuations are extracted and $C_n^2$ is derived. 

The use of recent high-density radiosonde profile at Trappes (France) and Hilo, HI (USA) showed the capability of the approach to obtain average $C_n^2$ profiles having similarities with well-known empirical profiles. This enabled to obtain site-dependent profiles using only classical radiosonde measurements, enabling further studies of geographical and seasonal variations of mean $C_n^2$ profiles. 
Furthermore, comparisons with thermosonde measurements showed good agreement and highlighted the possible identification of optical turbulent layers related to temperature inversions. 

The main limitations of the approach are related to the input meteorological data used. Indeed, the scale of the turbulent structures that can be modeled is directly related to the available radiosonde vertical resolution. Moreover, radiosonde data are only available at some places in the world and at particular instants. A possible solution to this limitation is the use of meteorological quantities coming directly from NWP simulations as inputs of the $C_n^2$ model. 

\section*{Data Availability Statement}
Radiosonde data used in this research are publicly available in the University of Wyoming (UWYO) Atmospheric Science Radiosonde Archive. They can be accessed at \url{http://weather.uwyo.edu/upperair/bufrraob.shtml}. Thermosonde measurements come from the T-REX campaign \cite{TREX} and are accessible using \url{https://doi.org/10.26023/6HCG-E9WC-H0E}.

\section*{Acknowledgment}
The University of Wyoming is thanked for the access to high-density radiosonde profiles through its website. AFRL and NCAR are thanked for the access to thermosonde data of the T-REX campaign. Data provided by NCAR/EOL under the sponsorship of the National Science Foundation, \url{https://data.eol.ucar.edu/}.

\appendix
\section{Influence of sampling distance on $C_n^2$ estimation}
\label{app:A}

For simplicity, a sketch of the proof is given in 1D. Assuming that the refractive index fluctuations $n_1(r)$ are sampled in the vertical direction, the equivalent sampled field $n_{1s}(r)$ is
\begin{align}
    n_{1s}(r)\:&=\:n_1(r)\sum_{k=-\infty}^\infty \delta(r-k \Delta z -z_0)\\
    &=\: \sum_{k=-\infty}^\infty \underbrace{n_1(k\Delta z + z_0)}_{n_1[k]} \delta(r-k \Delta z -z_0),
\end{align}
where $\delta(r)$ is the Dirac delta function, $\Delta z$ is the sampling distance and $z_0$ is the initial position where sampling is started ($z_0 \in [0,\Delta z[$). This initial position is assumed to be random in order to make the sampled refractive index field homogeneous \cite{leon1994probability}. The discrete field is denoted $n_1[k]$.

With these homogeneity and one-dimensional fluctuation hypotheses, the structure function in the inertial range is given by \cite{tatarskii1971effects}
\begin{align}
    D_n(\rho)\:&=\: \left \langle \left ( n_1(r+\rho)-n_1(r)\right)^2\right \rangle\\&=\:2\left(B_n(0)-B_n(\rho)\right)\:=\:C_n^2 ~\rho^{2/3},
    \label{eq:strucB}
\end{align}
where $B_n(\rho)$ is the covariance function \revtwo{which} definition is
\begin{align}
    B_n(\rho)\:=\: \left \langle n_1(r+\rho)~n_1(r)\right \rangle.
\end{align}
The covariance function of the sampled field is given by (\ref{eq:app1}). Assuming that $z_0$ is known and independent of $n_1(r)$, the expectation can first be computed on $n_1$ and (\ref{eq:app2}) is obtained, with $B_n[k-k']$ the discrete covariance function.
\begin{figure*}
\noindent\rule{\textwidth}{0.6pt}
\begin{align}
    B_{n_s}(\rho) \:&=\:\left \langle \left(\sum_{k=-\infty}^\infty n_1[k] \delta(r+\rho-k \Delta z -z_0)\right)\left(\sum_{k'=-\infty}^\infty n_1[k'] \delta(r-k' \Delta z -z_0)\right)\right \rangle,
    \label{eq:app1}
\end{align}
\begin{align}
    B_{n_s}(\rho|z_0) \:&=\: \sum_{k=-\infty}^\infty \sum_{k'=-\infty}^\infty \underbrace{\left \langle n_1[k]~n_1[k'] \right \rangle}_{B_n[k-k']} \delta(r+\rho-k \Delta z -z_0) ~\delta(r-k' \Delta z -z_0).
     \label{eq:app2}
\end{align}
\begin{align}
    B_{n_s}(\rho) \:&=\: \sum_{k=-\infty}^\infty \sum_{k'=-\infty}^\infty B_n[k-k'] \int_{0}^{\Delta z}\frac{1}{\Delta z} \delta(r+\rho-k \Delta z -z_0) ~\delta(r-k' \Delta z -z_0) ~\mathrm{d}z_0\nonumber \\
    &=\sum_{k=-\infty}^\infty \sum_{k'=-\infty}^\infty \frac{B_n[k-k']}{\Delta z} \int_{k'\Delta z-r}^{(k'+1)\Delta z-r} \delta(\rho-(k-k') \Delta z -s) ~\delta(s)~ \mathrm{d}s\nonumber\\
    &=\sum_{k=-\infty}^\infty  \frac{B_n[k-k'^*]}{\Delta z}~ \delta(\rho-(k-k'^*) \Delta z) \nonumber\\
    &= \sum_{i=-\infty}^{\infty}\frac{B_n[i]}{\Delta z}~ \delta(\rho-i \Delta z),
    \label{eq:app3}
\end{align}
\noindent\rule{\textwidth}{0.6pt}
\end{figure*}

 Then, the expectation on $z_0$ is taken, assuming $z_0$ is uniformly distributed between $0$ and $\Delta z$ such that its probability density function $T_{Z_0}(z_0)$ is
\begin{equation}
    T_{Z_0}(z_0) \:=\:\begin{cases} \frac{1}{\Delta z} \:\:\text{if}\:\: 0 \leq z_0 < \Delta z, \\ 0 \:\:\:\:\: \text{otherwise}.\end{cases}
\end{equation}
This leads to the integration presented in (\ref{eq:app3}), according to the definition of the expectation, and where the change of variable $s=k'\Delta z+z_0-r$ has been used, as well as the fact that there exists only one value of $k'$ (denoted $k'^*$) such that $s=0$ is in the integration domain and $\int_{k'^*\Delta z-r}^{(k'^*+1)\Delta z-r} \delta(s)~ \mathrm{d}s = 1$. Finally, the change of variable $i=k-k'^*$ is used.

In terms of structure function, (\ref{eq:strucB}) becomes
\begin{footnotesize}
\begin{align}
    D_{n_s}(\rho)\:&=\:2\left(\sum_{i=-\infty}^{\infty}\frac{B_n[i]}{\Delta z}~ \delta(i \Delta z)-\sum_{i'=-\infty}^{\infty}\frac{B_n[i']}{\Delta z}~ \delta(\rho-i' \Delta z)\right)\nonumber\\
    &=\: C_n^2 ~\rho^{2/3}.
\end{align}
\end{footnotesize}

Since $D_{n_s}(\rho)$ is only known for distances $\rho$ multiple of the sampling distance, i.e. $\rho=m\Delta z$ with $m\in \mathbb{Z}$:
\begin{footnotesize}
\begin{align}
    D_{n_s}(m\Delta z)\:&=\:2\left(\sum_{i=-\infty}^{\infty}\frac{B_n[i]}{\Delta z}\delta(i \Delta z)-\sum_{i'=-\infty}^{\infty}\frac{B_n[i']}{\Delta z}\delta((m-i') \Delta z)\right)\nonumber \\
    &=\: 2\left(\frac{B_n[0]}{\Delta z}-\frac{B_n[m]}{\Delta z}\right)\:=\:C_n^2 ~(m\Delta z)^{2/3},
    \label{eq:Dns}
\end{align}
\end{footnotesize}
using the properties of the delta functions.

In practice, the structure function is computed in the discrete domain, i.e.
\begin{align}
    D_{n}[m]\:&=\:\left \langle \left(n_1[k+m]-n_1[k]\right)^2 \right \rangle \nonumber \\
    &=\left \langle n_1^2[k+m] \right \rangle -2 \left \langle n_1[k+m]~n_1[k] \right \rangle + \left \langle n_1^2[k] \right \rangle \nonumber\\
    &=\: 2\left(B_n[0]-B_n[m]\right).
\end{align}
Using this last result and (\ref{eq:Dns}), our estimation of $C_n^2$ based on the discrete structure function is therefore
\begin{align}
    C_n^2\:=\:\frac{1}{\Delta z}\frac{D_n[m]}{(m\Delta z)^{2/3}}.
\end{align}
This corresponds to (\ref{eq:Cn2est}) where the ergodicity hypothesis is used to compute the structure function $D_n[m]$ using a spatial average on 3 points.

%\section*{References}
\vspace{-0.1cm}
\bibliographystyle{ieeetr}
\bibliography{biblio}
\vspace{-0.2cm}

\end{document}